\documentclass[12pt]{article}
\usepackage{amsmath}

\input BoxedEPS
\SetRokickiEPSFSpecial  %% for dvips by Tom Rokicki, for VMS
\HideDisplacementBoxes

\begin{document}

\title{Cold Trapped Atoms: A Mesoscopic System{\LARGE \thanks{Expanded version of an
invited talk given at the third joint meeting of Chinese physicists worldwide,
Chinese University of Hong Kong, August, 2000.}} }
\author{Kerson Huang\\
\small\it Department of Physics and Center for Theoretical
Physics\\
\small\it Massachusetts Institute of Technology\\
\small MIT-CTP\#3048}
\date{}
\maketitle

\begin{abstract}
The Bose-Einstein condensates recently created in trapped atomic gases are
mesoscopic systems, in two senses: (a) Their size fall between macroscopic and
microscopic systems; (b) They have a quantum phase that can be manipulated
in experiments. We review the theoretical and experimental facts about trapped
atomic gases, and give examples that emphasize their mesoscopic characters.
One is the dynamics of collapse of a condensate with attractive interactions.
The other is the creation of a 1D kink soliton that can be used as a
mode-locked atom laser.

\end{abstract}

\section{Mesoscopic systems}

A mesoscopic system has two distinctive characteristics:

\begin{itemize}
\item It contains many more particles than a microscopic system, but is small
enough for us to manipulate in the laboratory.

\item It possesses a quantum phase that we can control.
\end{itemize}

The term ``mesoscopic'' has been associated with ``quantum dots,'' in which
the passage of electrons can be controlled singly. Now we have new
mesoscopic systems, with the achievement of Bose-Einstein condensation in a
system of trapped atoms. These Bose condensates typically contain the order
of 10$^{6}$ atoms, in a magnetic or optical trap, at a temperature of the order
of 10$^{-7}$K. After a brief summary of known facts, we shall give examples
of both aspects mentioned above. For more details and literature, we refer to
recent review papers on experiments~\cite{review1} and theory~\cite{review2}.

\section{The Bose condensate}

As we know, there are two types of elementary particles: fermions of
half-integer spin, which cannot occupy the same state, and boson of integer
spin, which love to be in the same state. Bound states of elementary particles,
such as atoms and molecules, also fall into these categories. The bosonic atoms
we are dealing with include $^{87}$Rb, $^{23}$Na, $^{7}$Li, and atomic
hydrogen.

\begin{figure}[ht]
$$
\BoxedEPSF{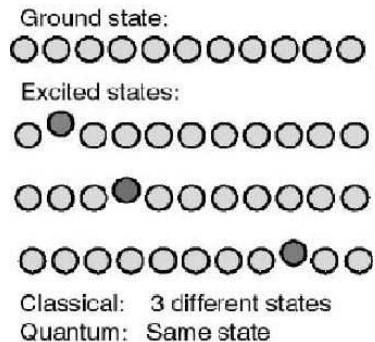 scaled 500}  %% scales it 50 percent
$$
\caption{The indistinguishability of particles reduces the number of excited
states, and makes the condensate stable.}
\label{f1}
\end{figure}

  In the ground states all bosonic atoms occupy the same state; forming the
Bose-Einstein condensate. What makes this an identifiable entity is its
stability, arising from the quantum-mechanical fact that atoms are
indistinguishable. In Fig.~\ref{f1} we illustrate how this reduces the number of
excited states. The situations illustrated would represent three different
states, if the particles were distinguishable. But for indistinguishable particles
they are one and same the same state. Because of this, there are fewer ways
to excite the condensate, and hence it has a high degree of stability.

In   fact, when the temperature is raised from absolute zero, there remains a
finite fraction of particles in the condensate. The condensate fraction
decreases with rising temperature, and disappears only above a finite
temperature $T_{c}$, the critical temperature of the Bose-Einstein
condensation.

In contrast, for ``Boltzmann statistics'', the ground state wave function is the
same, but there will be a high density of excited states. In this case, no
condensate exists at any finite temperature. That is, the transition
temperature is at absolute zero.

\begin{figure}[ht]
$$
\BoxedEPSF{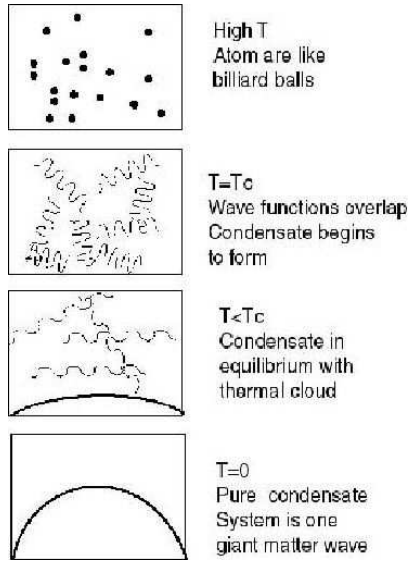 scaled 500}  %% scales it 50 percent
$$
\caption{The approach to Bose-Einstein condensation.}
\label{f2}
\end{figure}

The process of Bose-Einstein condensation is illustrated in Fig.~\ref{f2}.

\section{Experimental techniques}

\begin{figure}[ht]
$$
\BoxedEPSF{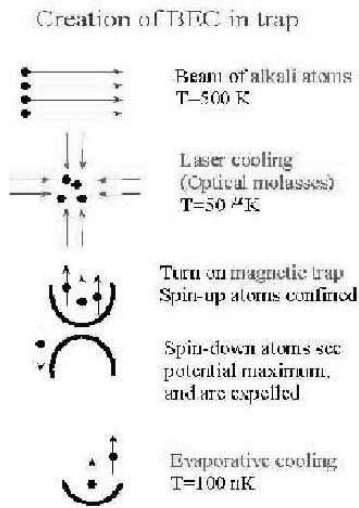 scaled 500}  %% scales it 50 percent
$$
\caption{Experimental steps in the trapping and cooling of atoms.}
\label{f3}
\end{figure}

The process to achieve BEC in the laboratory is illustrated in Fig.~\ref{f3}. An
atomic beam was slowed down by laser cooling, and a magnetic trap was
turned on.
The spin up atoms are confined, while the spin down one are expelled. Final
cooling is achieved through evaporation and rethermalization. To evaporate
the fast atoms, rf light of appropriate frequency is turned on to induce spin
flip. In the final condensate, we have typically
\begin{align} N  &  \sim10^{6}\nonumber\\ V  &  \sim10^{-6}\text{
cm}^{3}\nonumber\\
\frac{N}{V}  &  \sim10^{12}\text{ cm}^{-3}
\end{align} The density is some seven orders of magnitude smaller than that
of air in the atmosphere.

\begin{figure}[ht]
$$
\BoxedEPSF{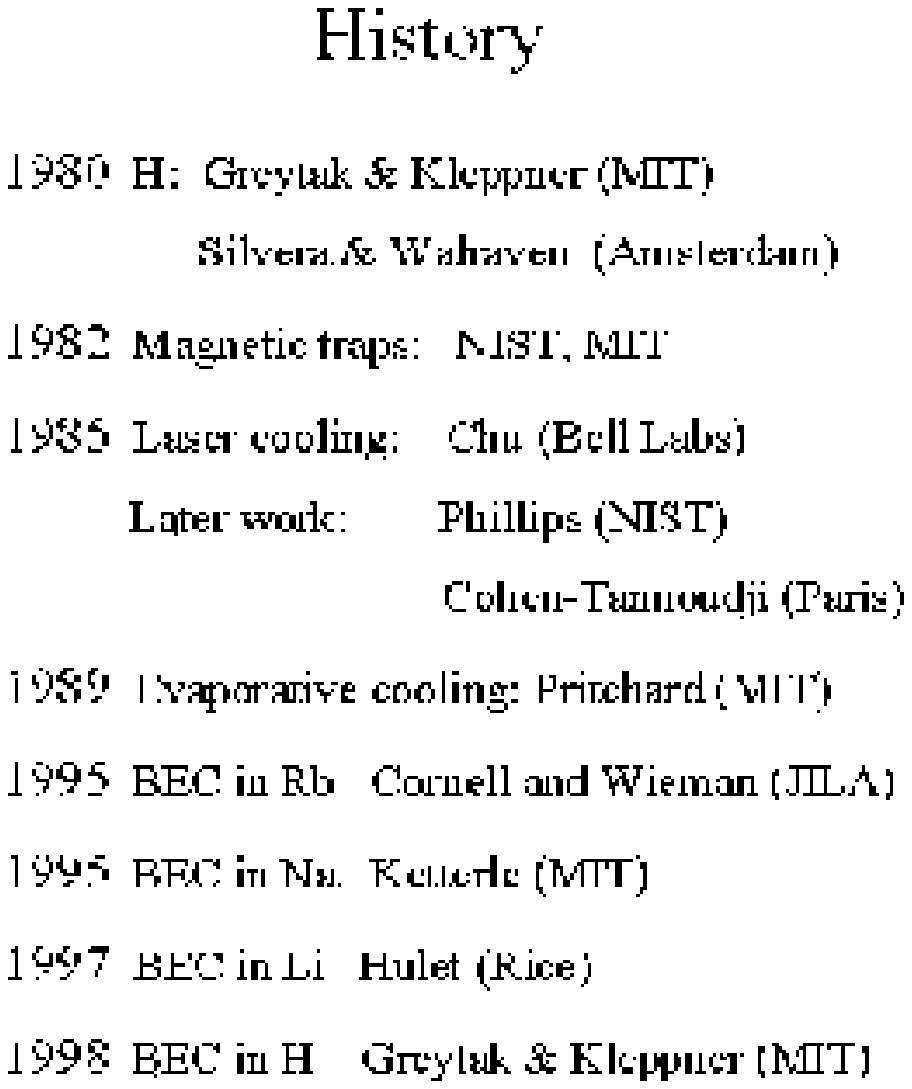 scaled 600}  %% scales it 50 percent
$$
\caption{History of atom trapping and cooling.}
\label{f4}
\end{figure}

A brief history of Bose-Einstein condensation in trapped gases is given
Fig.~\ref{f4}.

\section{Helium clusters}

Long before trapped atoms, of course, Bose-Einstein condensation had been
seen, though indirectly, in the superfluidity of liquid $^{4}$He.
Andronikashvili performed the classic experiment on the superfluid fraction,
by measuring the moment of inertia of a stack of rotating plates immersed in
liquid helium, as a function of temperature. The idea was that the normal
fluid would be dragged into rotation with the plates, but the superfluid
component would not participate in the motion.

The ultimate miniaturization of the Andronikashvili experiment has now been
done~\cite{clusters} in $^{4}$He clusters of a few hundred atoms, with the
rotating plates replaced by a rotating atom of $^{3}$He. Superfluidity was
reported in a cluster with 60 atoms. This is a very interesting mesoscopic
system, but beyond the scope of our discussion here.

\section{Quantum coherence}

For an ideal Bose gas, the condensate is made up of particles in the same
single-particle quantum state. When there are interparticle interactions,
single-particle states are no longer meaningful; but we can define the
condensate wave function as the quantum amplitude for removing a particle
from the ground state:
\begin{equation}
\psi(\mathbf{r},t)=\sqrt{n_{s}}e^{i\phi}
\end{equation} 
where $n_{s}$ is the condensate density, and $\phi$ is the
quantum phase of the condensate.

\begin{figure}[ht]
$$
\BoxedEPSF{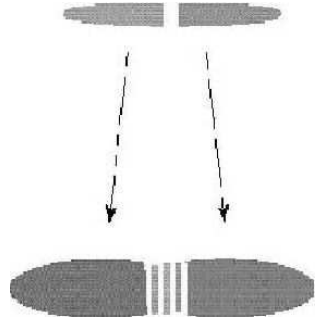 scaled 500}  %% scales it 50 percent
$$
\caption{ A condensate in free fall is cut by a laser knife, and the two halves
continue to fall, expand, and overlap.}
\label{f5}
\end{figure}

\begin{figure}[ht]
$$
\BoxedEPSF{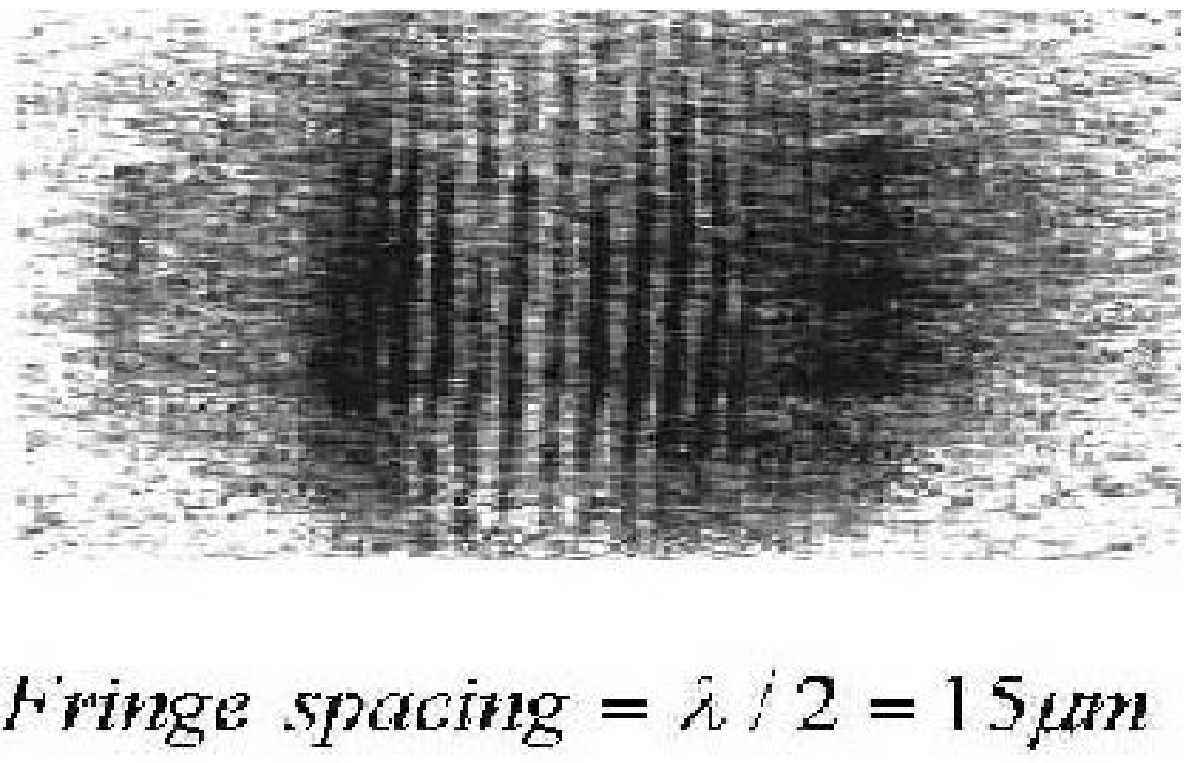 scaled 500}  %% scales it 50 percent
$$
\caption{Interference fringes in two overlapping condensates of sodium 
atoms. (\emph{Courtesy W. Ketterle.})}
\label{f6}
\end{figure}

The quantum coherence of the condensate is a remarkable thing. In
experimental traps, the condensate can have a spatial extension as large as
$1$ mm, which is almost visible to the naked eye, and yet this glob of matter
is characterized by a single wave function, with a single phase. This has been
demonstrated experimentally by switching off the magnetic trap, cutting the
condensate in half with a laser ``knife,'' and allowing the two halves to fall
under gravity. The two condensates expand as they drop, and eventual
overlap in space, creating a interference pattern, as illustrated in Fig.~\ref{f5}.
An actual photograph of the interference fringes in shown in Fig.~\ref{f6}.

We can describe two non-overlapping free condensates by wave functions that
are nearly plane waves:
\begin{align}
\psi_{1}  &  =C_{1}e^{i\left(  \mathbf{k}_{1}\mathbf{\cdot}r-\omega
_{1}t\right)  }\nonumber\\
\psi_{2}  &  =C_{2}e^{i\left(  \mathbf{k}_{2}\mathbf{\cdot r}-\omega
_{2}t\right)  }
\end{align} When they overlap, they become one condensate with wave
function
\begin{equation}
\psi=\psi_{1}+\psi_{2}
\end{equation} 
The density of the system is given by
\begin{equation} |\psi|^{2}=|C_{1}|^{2}+|C_{1}|^{2}-2\text{Re}\left[  C_{1}^{\ast}
C_{2}e^{i\left(  \mathbf{k}_{1}\mathbf{-k}_{2})\cdot\mathbf{r}-i(\omega
_{1}-\omega_{2})t\right)  }\right]
\end{equation} where the last term exhibits the interference fringes. From
Fig.~\ref{f5}, the fringe spacing gives a half wavelength $\lambda/2=$15
$\mu$m, which corresponds to a relative wave number
\begin{equation} k_{1}-k_{2}=\frac{2\pi}{\lambda}
\end{equation} The relative velocity of the two condensates was therefore
given by
\begin{equation} v_{1}-v_{2}=\frac{\hbar}{m}\left(  k_{1}-k_{2}\right) 
=0.05\text{ cm/s}
\end{equation}

\begin{figure}[ht]
$$
\BoxedEPSF{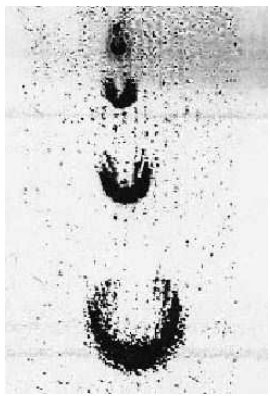 scaled 500}  %% scales it 50 percent
$$
\caption{The first atom laser. (\emph{Courtesy W.
Ketterle.})}
\label{f7}
\end{figure}

Atoms in the condensate in a magnetic trap can be ejected by inducing spin
flips, through application of rf light. The expelled atoms fall under gravity as
pulses of coherent matter, as shown in Fig.~\ref{f7}. This is the first example
of an atom laser~\cite{laser1}.

\section{Nonlinear Schr\"{o}dinger equation}

The condensate wave function satisfies a nonlinear Schr\"{o}dinger equation
(NLSE), also called the Gross-Pitaveskii equation, in a mean-field
approximation that neglects the coupling to the thermal cloud:
\begin{equation} i\hbar\frac{\partial\psi}{\partial t}=\left( 
-\frac{\hbar^{2}}{2m}\nabla ^{2}+V(r)+g|\psi|^{2}\right)  \psi\label{NLSE}
\end{equation} with
\begin{equation} g=\frac{4\pi\hbar^{2}a}{m}
\end{equation} where $a$ is the S-wave scattering length. The normalization
condition
\begin{equation}
\int d^{3}r|\psi|^{2}=N
\end{equation} determines the number of particles in the condensate. This is
a constant of the motion, but its value cannot be changed by convention,
contrary to that for an ordinary wave function, because the equation is
nonlinear.

\section{The scattering length}

\begin{figure}[ht]
$$
\BoxedEPSF{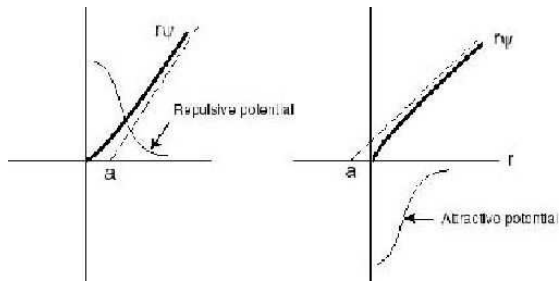 scaled 500}  %% scales it 50 percent
$$
\caption{Scattering length a for potentials without bound states.}
\label{f8}
\end{figure}

The scattering length $a$ is illustrated in Fig.~\ref{f8}. The reduced relative
wave function of two atoms, as a function of the separation $r$, approaches a
linear asymptote beyond the range of the potential. The scattering length is
the the intercept of the asymptote on the $r$ axis.

\begin{figure}[ht]
$$
\BoxedEPSF{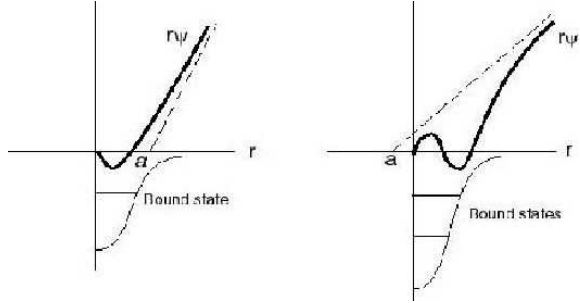 scaled 500}  %% scales it 50 percent
$$
\caption{Scattering length a for potentials without bound states.}
\label{f9}
\end{figure}

When the potential has no bound state, as illustrated in Fig.~\ref{f9}, a
repulsive potential give $a>0$, and an attractive potential give $a<0$. When
there are bound states in a attractive potential, however, the scattering length
can have either sign. As the potential gets more attractive, it changes sign
whenever a bound state occurs.

One of the exciting recent developments is that the scattering length can be
tuned, from $-\infty$ to $\infty$, by changing an external magnetic field,
through the Feshbach resonance in atomic scattering, which is similar to a
bound state~\cite{Feshbach}. This realizes the usual thought experiment of
changing the coupling constant in the Hamiltonian.

It should be emphasized that $a$ is a parameter that describes scattering in
3D in the limit of zero energy, where the shape of the potential is irrelevant.
At higher energies, the shape of the potential does make a difference, and one
must include other parameters, such as the effective
range~\cite{StatMech}. 

\section{Quantum tunneling}

The trapped atoms that form a Bose condensate at such low temperatures
would have solidified in free space. In the trap, the zero-point motion kept
them apart. In the case of $^{6}$Li, which has negative scattering length, the
condensate is metastable, and decays via quantum tunneling, if the total
number of atoms $N$ is sufficiently small. If $N$ exceeded a critical number,
however, the system collapses~\cite{Ueda,Eleftheriou}. Let us discuss
each of these scenarios.

For negative scattering length,  we have $g<0$. Let us denote
\begin{equation} G=-g=-\frac{4\pi\hbar^{2}a}{m}
\end{equation} We can see qualitatively that in free space the condensate
collapses in spatial dimension $D>2$. The energy density of system is
\begin{equation} E=\frac{\hbar^{2}}{2m}|\nabla\psi|^{2}-G|\psi|^{4}
\end{equation} Suppose the system is confined to a region of dimension $R$.
Then $\psi |^{2}\sim N/R^{D}$, so
\begin{equation}
E\sim\frac{\hbar^{2}}{2m}\frac{N}{R^{D+2}}-\frac{gN^{2}}{R^{2D}}
\end{equation} Thus $E\rightarrow-\infty$ when $R\rightarrow0$, if $D>2$.

If there is a confining potential, then collapse happens only if $N$ is
sufficiently large, because there is a potential barrier against collapse. To see
this, consider the Hamiltonian corresponding to (\ref{NLSE}):
\begin{equation} H[\psi]=\int d^{3}r\left[ 
-\frac{\hbar^{2}}{2m}\psi^{\ast}\nabla^{2}
\psi+V(r)\psi^{\ast}\psi-\frac{G}{2}(\psi^{\ast}\psi)^{2}\right]
\end{equation} and choose a harmonic oscillator potential
\begin{equation} V(r)=\frac{1}{2}m\omega^{2}r^{2}
\end{equation}

\begin{figure}[ht]
$$
\BoxedEPSF{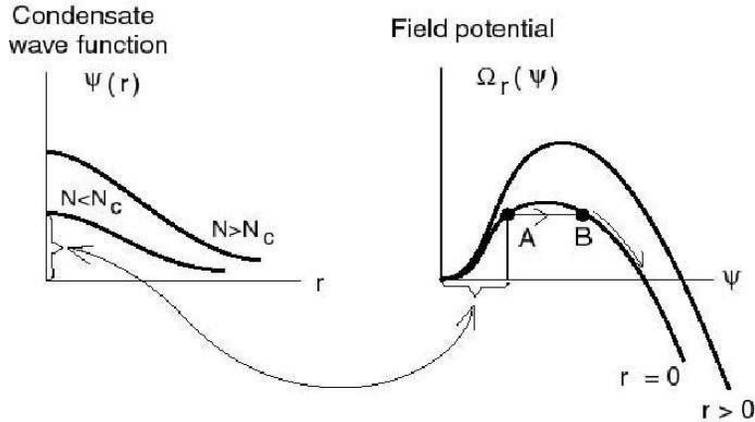 scaled 500}  %% scales it 50 percent
$$
\caption{The field potential as seen by the condensate wave function. Local
quantum tunneling $A\rightarrow B$ occurs across an energy barrier. If the
wave function at $r$ goes over the energy barrier, then there is local collapse
at that point.}
\label{f10}
\end{figure}

For qualitative arguments, let us consider a confined wave function of a
Gaussian shape
\begin{equation}
\psi_{0}(r)=C_{0}\exp\left(  -\alpha r^{2}/d_{0}^{2}\right)
\end{equation} where
\begin{equation} d_{0}=\sqrt{\frac{\hbar}{m\omega}}
\end{equation} The parameter $\alpha$ describes the width of the Gaussian.
A Gaussian narrower than the harmonic oscillator wave function corresponds
to $\alpha>1$. We can then replace
$-\frac{\hbar^{2}}{2m}\nabla^{2}\psi\rightarrow\chi (r)\psi_{0}$, where
\begin{equation}
\chi(r)=\frac{\hbar^{2}}{2md_{0}^{2}}\left[  3\alpha-\left(  \frac{r}{d_{0}
}\right)  ^{2}\alpha^{2}\right]
\end{equation} Thus the Hamiltonian can be represented in the form
\begin{equation} H[\psi]=\int d^{3}r\Omega_{r}(\psi)
\end{equation} where the ``field potential'' $\Omega_{r}(\psi)$ is given by
\begin{equation}
\Omega_{r}(\psi)=\left[  V(r)+\chi(r)\right]  |\psi|^{2}-\frac{G}{2}|\psi|^{4}
\end{equation} This potential is plotted as a function of
$|\psi|$ for different $rT,$ in the right panel of Fig.~\ref{f10}. There is an
energy barrier with height
\begin{equation} W_{r}=\frac{1}{G}\left[  V(r)+\chi(r)\right]  ^{2}
\end{equation} which is nonzero at $r=0$, because $\chi(0)\neq0$. At large
$|\psi|$ the field potential tends to $-\infty$. In reality, of course, the NLSE
ceases to be valid somewhere along the drop, for other physical effects, such
as solidification, come in.

A typical initial wave function $\psi_{0}(r)$ is shown on the left panel of
Fig.~\ref{f10}. At a given $r$, we can measure the wave function on the left
panel, and transfer it to the horizontal axis on the right panel. If it lands to
the left of the barrier maximum for the particular $r,$ then the system at that
point is classically stable, but can decay via quantum tunneling as indicated
by the classically forbidden path $A\rightarrow B$. If it lands to the right, the
system at that $r$ will rapidly collapse to a state of high density.

Since $\psi_{0}(r)$ has a maximum at $r=0$ with value proportional to
$\sqrt{N}$, the system will decays local tunneling, if it is classically stable at
$r=0$. Otherwise, the system in a region about $r=0$ will rapidly collapse. The
condition for stability against collapse is therefore
$\Omega_{0}(\psi_{0}(0))<W_{0}.$ Within the Gaussian assumption, this
corresponds to $N<N_{c}$, with
\begin{equation}
N_{c}=\frac{3}{8}\sqrt{\frac{\pi}{\alpha}}\frac{d_{0}}{|a|}=\frac{0.665}
{\sqrt{\alpha}}\frac{d_{0}}{|a|} \label{Nc}
\end{equation} In the experiments with $^{7}$Li at Rice University~\cite{Rice},
the parameters are
\begin{align} a  &  =-1.45\text{ nm}\nonumber\\ d_{0}  &  =3\text{
m}\nonumber\\
\frac{2\pi}{\omega}  &  =6000s \label{param}
\end{align} With the choice $\alpha=1.8$ (See later), we obtain
$N_{c}\approx1200$ from (\ref{Nc}).

For $N<N_{c}$, the condensate decays slowly via quantum tunneling, as
indicate by the path $A\rightarrow B$ in Fig.~\ref{f10}. Since the tunneling
path depends on
$r$, the tunneling is a local phenomenon, proceeding at different rates at
different distances from the center of the trap.

It should be noted that quantum tunneling is a phenomenon not included in
the NLSE, which is a classical approximation. The tunneling probability can be
calculated via a Feynman path integral, and in a WKB-type approximation can
be obtained by solving the NLSE in imaginary time~\cite{Ueda}.

\begin{figure}[ht]
$$
\BoxedEPSF{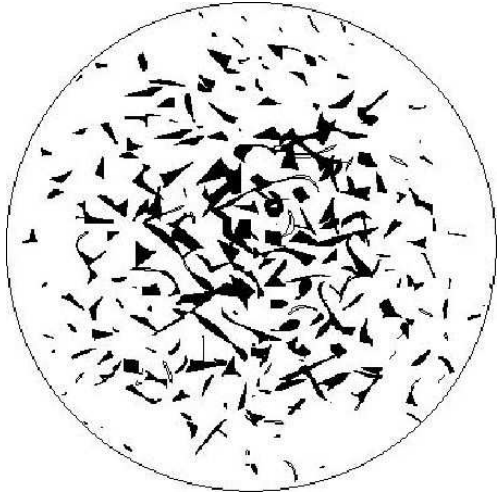 scaled 500}  %% scales it 50 percent
$$
\caption{High-density droplets  ``ooze out'' of the condensate via local
quantum tunneling.}
\label{f11}
\end{figure}

Qualitatively, we can describe the phenomenon as follows. The condensate
locally makes a quantum transition from a low-density state $A$ to a
high-density state $B$. The time at which this happens is random, and follows
a statistical distribution. Since the total number of particles is conserved,
what we see is that high-density droplets would ``ooze out'' of the condensate,
as illustrated in Fig.~\ref{f11}. Once form, the droplets will tend to contract
to states of even higher density, and will leave the trap through effects not
included in our Hamiltonian, such as spin-flip scattering.

A more rigorous treatment of the field potential, independent of the Gaussian
assumption, may be modeled after that of the ``effective potential'' in
quantum field theory~\cite{QFT}.

It should be noted that, unless $N\approx N_{c}$, the tunneling probability is
very small. In reality, it would be masked by the decay of the condensate
through spin-flip scattering, an effect that we have not taken into account.

\section{Collapse of the condensate}

When $\dot{N}>N_{c}$, the condensate with negative scattering length
collapses. As we can see from Fig.~\ref{f10}, the collapse does not happen all
at once over the entire condensate, but begins at the center of the condensate,
where the density is highest. A black hole opens up at the center, in which,
theoretically according to the NLSE, the density fluctuations becomes infinite,
and the pressure becomes negative~\cite{Ueda}.

Numerical solutions of the NLSE have been obtained~\cite{Eleftheriou}, with the
perimeters (\ref{param}). After the condensate was created in some arbitrary
state, the wave function quickly adjusted itself to a Gaussian-like
quasi-stationary state, with a width corresponding to $\alpha=1.8$. This
situation persists infinitely if $N$ is sufficiently small.

\begin{figure}[hbt]
$$
\BoxedEPSF{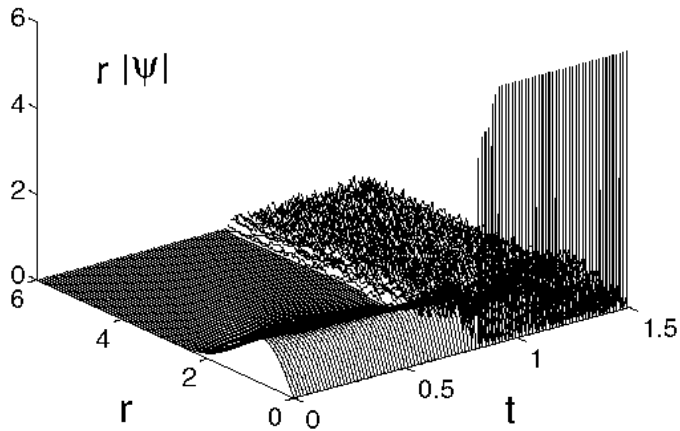 scaled 500}  %% scales it 50 percent
$$
\caption{The condensate is stationary for a time, and suddenly collapses locally,
near the center of the trap, opening up a black hole. The density in the black
hole shoots up as time goes on, fed by waves of implosion, which make the
density look turbulent. The density height is truncated because of finite grid
size in the numerical simulation.}
\label{f12}
\end{figure}

If $N$ is larger than a critical value, then the quasi-stationary states lasts for
a time, but then suddenly a finite volume about the center implodes, creating
a black hole. Thereafter, waves of implosion arrives from all over the
condensate, whose density now appears to be turbulent. These features are
summarized in Fig.~\ref{f12}, where $r\vert \psi|$ is plotted as a function of
$r$ in unit of $d_{0}$, and $t$ in unit of 2$/\omega$, for a fixed $N$. By
decreasing $N$ until the time of collapse goes to infinity, it was found that
$N_{c}\approx1200.$

\begin{figure}[ht]
$$
\BoxedEPSF{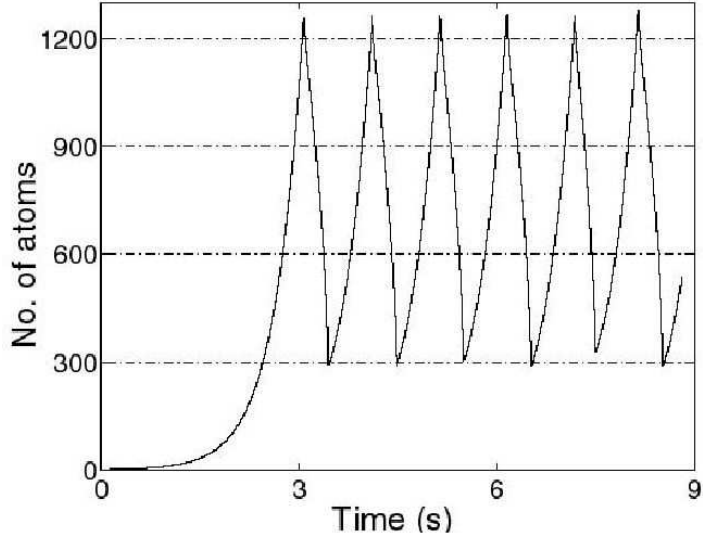 scaled 500}  %% scales it 50 percent
$$
\caption{The condensate goes through steady-state growth-collapse cycles, in the
presence of gain-loss mechanisms.}
\label{f13}
\end{figure}

In the experimental situation, the condensate in the trap can exchange atoms
with an uncondensed thermal cloud, which contains far more atoms than the
condensate. Equilibrium is reached when the chemical potentials equalize. The
atoms can be depleted through two and three-body spin-flip collisions, and
the latter becomes especially important where the density becomes large.
These gain and loss mechanisms can be simulated by adding extra
pure-imaginary terms to the NLSE, with a linear gain term, and a loss term
proportional to 
$\vert \psi|^{4}\psi.$ With these additions, the system settles down to steady
state in which $N$ oscillates in growth-collapse cycles, as shown in
Fig.~\ref{f13}. Again, we obtain $N_{c}\approx1200$.

\section{Solitons in a condensate ring}

\bigskip A spatial variation of the phase $\phi$ of the condensate wave
function gives rise to a superfluid velocity
\begin{equation}
\mathbf{v}_{s}=\frac{\hbar}{m}\nabla\phi
\end{equation} and a supercurrent density
\begin{equation} j_{s}=n_{s}v_{s}
\end{equation} Thus, flow patterns in the condensate can be created by
``phase engineering.'' This has been done to create
vortices~\cite{vortex1,vortex2} and solitons~\cite{Denschlag}.

\begin{figure}[ht]
$$
\BoxedEPSF{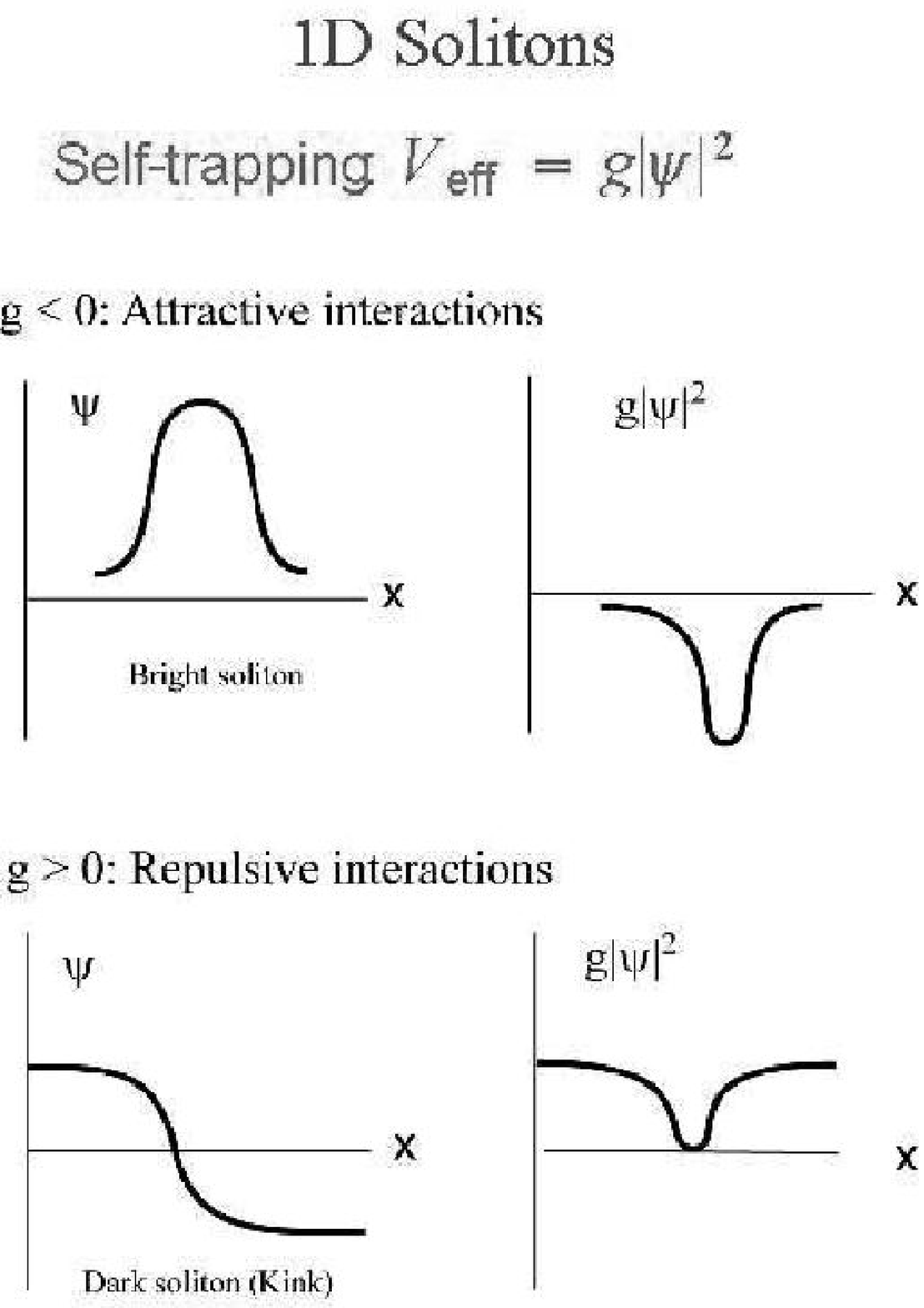 scaled 500}  %% scales it 50 percent
$$
\caption{Formation of 1D solitons via ``self-trapping''.}
\label{f14}
\end{figure}

One-dimensional solitons have been used in optical fibers for signal
transmission over thousands of kilometers, with minimal loss. The
electromagnetic field in an optical fiber obeys a NLSE of the same form as
(\ref{NLSE})~\cite{optical}. Thus, it is natural to wonder whether we can create
solitons in a 1D condensate. As illustrated in Fig.~\ref{f14}, the nonlinear term
$g|\psi|^{2}$ represents an effective potential, which is attractive for
$g<0$, and repulsive for $g>0$. $g<0$As we saw earlier, the case does not lead
to collapse in 1D. In this case, a localized wave packet created a attractive
well of the same shape, hence it would trap itself into a bright soliton. In the
case $g>0$, on the other hand, a ``kink'' configuration would self-trap, giving
rise to a dark soliton.

Consider now a ring-shaped condensate. If the radius of the ring is much
larger than the radius of the cross section, then we can regard the system as
one-dimensional, in the sense that transverse excitations have a much high
energy than the logitudinal ones, and may be ignored at low temperatures.
However, the cross sectional radius may still be large compared to the
scattering length, so that the system is still 3D with respect to interatomic
scattering.

We concentrate on the repulsive case with $g>0$. The solution to the NLSE
must satisfy the boundary condition 
\begin{equation}
\psi(\theta+2\pi,t)=\psi(\theta,t)\,
\end{equation} where $\theta$ is the angle around the ring. Let us write 
\begin{equation}
\psi(\theta,t)\,=f(\theta,t)\,e^{i\phi(\theta,t)\,}
\end{equation} For a kink soliton, $f$ must have a zero, and so changes sign
when $\theta$ increases by 2$\pi$. To satisfy the boundary condition, the
phase must change by an odd multiple of $\pi$ in one complete revolution.
The simplest case corresponds to a change of $\pi$ in the phase. This means
that the lowest kink soliton has angular momentum per particle $\hbar/2,$
for the same mathematical reason that an electron has spin 1/2.

The solution can be obtained analytically~\cite{Drummond}. The phase changes
most rapidly at the location of the kink, and its slope there gives the kink a
nonzero propagation velocity. Thus, the soliton is manifested as a dark spot
moving around the ring with a characteristic velocity. The width of the soliton
turns out to be proportional to $N^{-1/2}$, where $N$ is the number of atoms
in the condensate.

\section{A mode-locked atom laser}

The first atom laser illustrated in Fig.~\ref{f7} consists of pulses of coherent
atoms. However, the different pulses do not have definite phase with respect
to each other. There is an advantage to have a mode-locked laser --- one in
which the output pulses are phase coherent. In more recent experiments, a
steady stream of output pulses has been achieved~\cite{laser2}, but it is not
clear whether the pulses are coherent with respect to each other.

\begin{figure}[ht]
$$
\BoxedEPSF{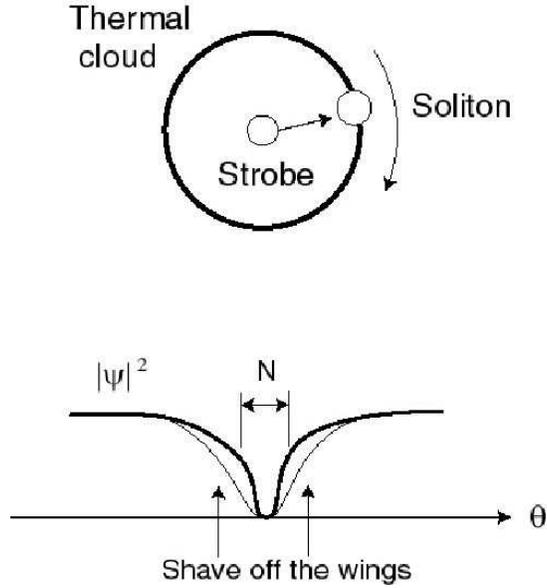 scaled 500}  %% scales it 50 percent
$$
\caption{Mode-locked atom laser created by stroboscopically shaving off
atoms at the kink of the dark soliton. The width grows back due to gain from
the thermal cloud.}
\label{f15}
\end{figure}

We can use of our kink soliton to make a mode-locked atom
laser~\cite{Drummond}. The physics is illustrated in Fig.~\ref{f15}. The ring
of condensate is surround by a thermal cloud, which tends to maintain an
equilibrium number of particles
$N$ in the ring. Now suppose we use some stroboscopic device to take
particles out the soliton, by enlarging its width, which is proportional to
$N^{-1/2}$. Then particles will flow from the thermal cloud to the condensate
to restore
$N$, and thus restoring the width.

An important question is how to create the kink soliton, and whether it is
robust against perturbations. In a simulation, the soliton was created by
applying a moving momentary repulsive potential, such as that created by a
laser beam, to create a hole in the condensate that moves at the characteristic
soliton speed. This hole develops into the kink soliton, plus all kinds of
excitations such as sound waves. To clean up the configuration, a periodic loss
mechanism was introduced, in the form of am imaginary stroboscopic
potential that also acts as outcoupler. At the same time, a gain mechanism
was put in to simulate the thermal cloud. All unwanted excitations rapidly
disappeared, leaving the kink soliton in steady-state equilibrium. The
stability of the soliton come from topology --- the fact that it has half-integer
angular momentum per particle.
\begin{figure}[ht]
$$
\BoxedEPSF{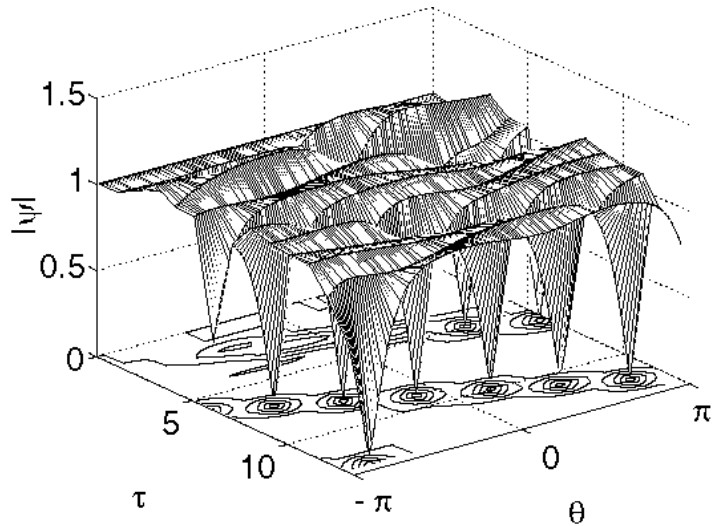 scaled 500}  %% scales it 50 percent
$$
\caption{The kink soliton was created by stirring the condensate at its
characteristic velocity, Other excitations also created in the stirring were
cleaned up by a stroboscopic loss mechanism, which also serves as output
coupler for the atom laser. The kink soliton survives because of topological
stability.}
\label{f16}
\end{figure}
\begin{figure}[ht]
$$
\BoxedEPSF{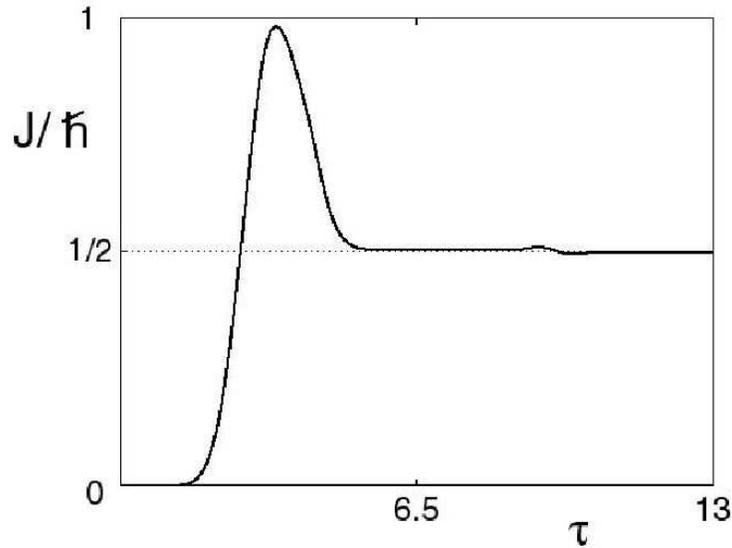 scaled 500}  %% scales it 50 percent
$$
\caption{The angular momentum per particle as function of time. After
transients die out, it approaches the half-integer value of a kink
soliton.}
\label{f17}
\end{figure}

\begin{figure}[ht]
$$
\BoxedEPSF{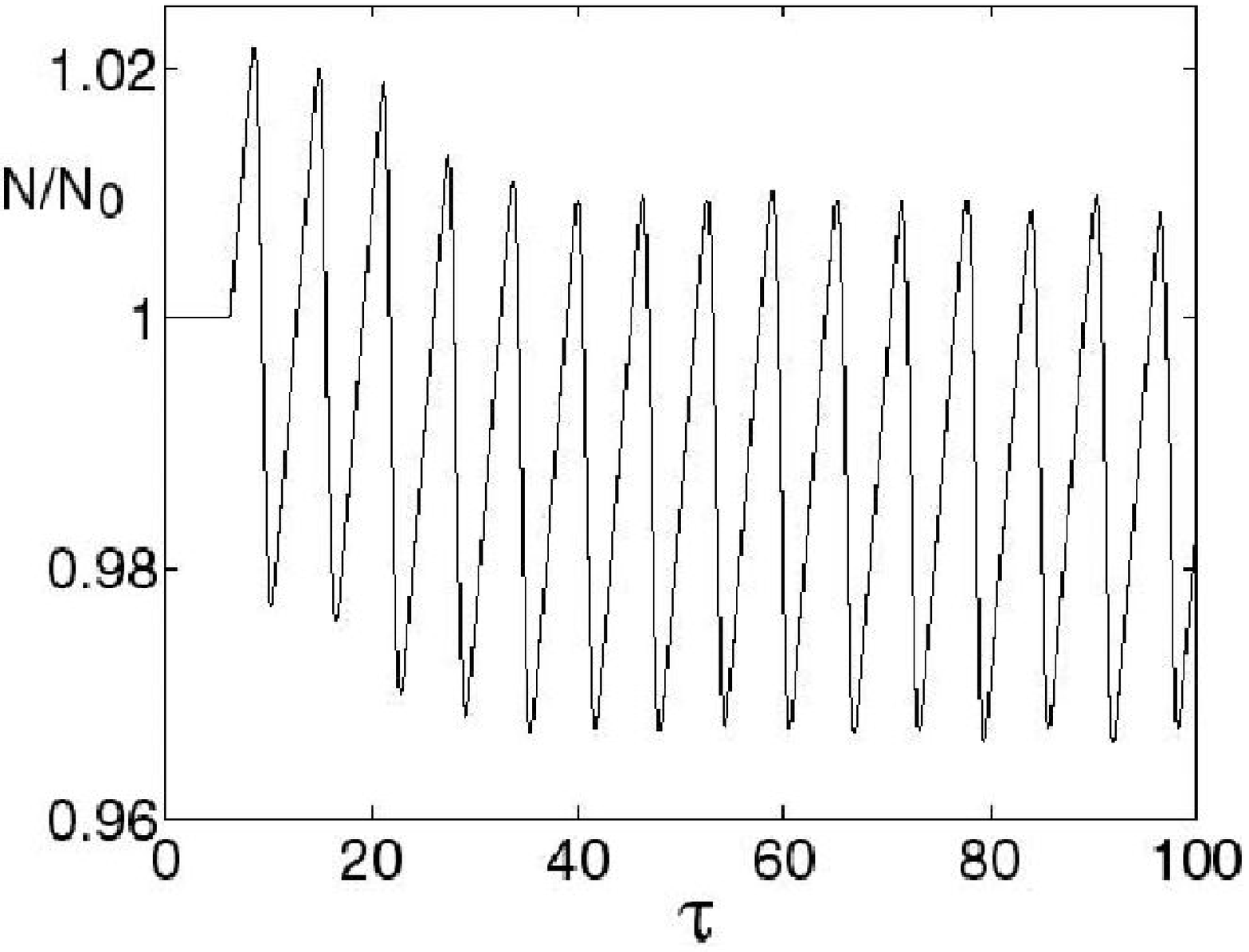 scaled 500}  %% scales it 50 percent
$$
\caption{Number of atoms in the ring condensate exhibits steady-state coherent
oscillations.}
\label{f18}
\end{figure}

The creation and stabilization of the soliton is depicted in Fig.~\ref{f16}, where
$|\psi|$ is plotted as a function of $\theta$ and $t$, with a contour plot
projected onto the $\theta$-$t$ plane. The fact that we have in fact a kink
soliton is indicted in Fig.~\ref{f17}, where the angular momentum per particle
is plotted as a function of time. Finally, Fig.~\ref{f18} shows the oscillation of
the number of particles in the ring. The particles lost from the ring form the
pulses in the mode-locked laser. They are coherent because the stroboscopic
action is coherent.

\subsection*{Acknowledgment} This work is support in part by a
U.S. Department of Energy cooperative agreement DE-FC02-94ER40818.


\begin{thebibliography}{9}                                                                                                

\bibitem {review1}W. Ketterle, D.S. Durfee, and D.M. Stumper-Kurn, in
\textit{Proceedings.of Internaltional School of Physics, Enrico Fermi Course
CXL}, M.\ Ignaxcio, S. Strinagari, and C.E. Wieman, Eds. (Soc. Italiana di
Fisica, Bologna, Italy, 1999).

\bibitem {review2}F. Dalfovo, S. Giorgini, L.P. Pitaevskii, and S. Stringari,
Rev. Mod. Phys. \textbf{71}, 463 (1999).

\bibitem {clusters}S. Grebenev, J.P. Toennies, and A. F. Vilesov, Science,
\textbf{279}, 2083 (1998).

\bibitem {laser1}M.-O. Mewes et. al., Phys. Rev. Lett. \textbf{78}, 582 (1977).

\bibitem {Feshbach}S. Inouye et. al., Nature, 392, 151 (1998);

\bibitem {StatMech}See K. Huang, \textit{Statistical Mechanics}, 2nd ed.
(Wiley, New York, 1987)., Sec.10.5.

\bibitem {Ueda}M. Ueda and K. Huang, Phys. Rev. A \textbf{60}, 3317 (1999).

\bibitem {Eleftheriou}A. Eleftheriou and K. Huang, Phys. Rev. A \textbf{61},
043601 (2000)..

\bibitem {Rice}C. C. Bradley, C. A. Sackett, and R. G. Hulet, \textit{Phys.
Rev. Lett}. \textbf{78}, 985 (1997); C. A. Sackett, H. T. C. Stoof, and R. G.
Hulet, \textit{Phys. Rev. Lett}. \textbf{80}, 2031 (1998).

\bibitem {QFT}See K. Huang, \textit{Quantum Field Theory,} (Wiley, New York,
1998), Sec.15.7.

\bibitem {vortex1}J.E. Williams and M.J. Holland, Nature, \textbf{401}, 568 (1999).

\bibitem {vortex2}M.R. Mathews et. al., Phys. Rev. Lett. \textbf{83}, 2498 (1999).

\bibitem {Denschlag}J. Denschlag \textit{et. al}.. Science, \textbf{287}, 97 (2000).

\bibitem {optical}A. Hasegawa, \textit{Optical Solitons in Fibers}, 2nd ed.
(Springer-Verlag, Berlin, 1995).

\bibitem {laser2}B.P. Anderson and M.A. Kasevich, Science,\textbf{27}, 1686
(1998); I Bloch, T.W. Haensch, and T. Esslinger, Phys. Rev. Lett. \textbf{82},
3008 (1999); M Kozuma et. al. Sience, 286, 2309 (1999).

\bibitem {Drummond}P.D. Drummond, A. Eleftheriou, K. Huang, and K.V.
Kheruntsyan, cond-mat/0002389.
\end{thebibliography}
\end{document}